\documentstyle[epsfig]{mn2e}

\begin{document}

\title[Prolate halos with radially varying eccentricity]{Dynamics of prolate spheroidal mass distributions with varying eccentricity}
\author[Rathulnath \& Jog]
{R. Rathulnath \thanks{E-mail : rathul@physics.iisc.ernet.in}
 , and 
 Chanda J. Jog \thanks{E-mail : cjjog@physics.iisc.ernet.in}\\
   Department of Physics,
Indian Institute of Science, Bangalore 560012, India \\
} 

\maketitle

\begin{abstract}
In this paper we calculate the potential 
 for a prolate spheroidal distribution as in a dark matter halo with a 
radially varying eccentricity.
This is obtained by summing up the shell-by-shell contributions of isodensity surfaces, which
are taken to be concentric and with a common polar axis and with an axis ratio that varies with radius.  Interestingly, 
the constancy of potential inside a shell
is shown to be a good approximation even when the isodensity contours are dissimilar spheroids, as long as the radial variation in eccentricity is small as seen in realistic systems.
 We consider three cases where the isodensity contours 
are more prolate at large radii, or are less prolate, or have a constant eccentricity. 
Other relevant physical quantities like the rotation velocity, the net orbital and vertical frequency due to the halo and an exponential disc of finite thickness embedded in it are obtained. We apply this to the kinematical origin
of Galactic
warp, and show that a prolate shaped halo is not conducive to making long-lived warps - contrary to what has been
proposed in the literature.
The results for a prolate mass distribution with a variable axis ratio obtained are general, and can be applied
 to other astrophysical systems such as prolate bars, for a more realistic treatment.
 
\end{abstract}

\begin{keywords}
galaxies: halos --- galaxies: kinematics and dynamics --- galaxies: structure --- methods: analytical
\end{keywords}

\section{Introduction}
A galactic disc is believed to be 
 embedded in a dark matter halo, but the actual shape of the dark matter halo is not yet well-understood. Usually, the halo is taken to be spherical 
for simplicity, or oblate. When a deviation from sphericity is considered at all, typically an oblate shape is used (e.g., Sackett \& Sparke 1990). 
A prolate halo has been used in a few cases, e.g. to study warps in galaxies (Sparke 1994, Ideta et al. 2000), and to explain the tidal streaming in the Galaxy (Helmi 2004).
In these the axis ratio is taken to be constant, again for simplicity.

Recently, it was shown that the
dark matter halo in the Milky Way is prolate with the
vertical-to-planar axis ratio monotonically increasing to 2.0
at 24 kpc (Banerjee \& Jog 2011). This was obtained by modeling 
the observed steeply flaring atomic hydrogen gas layer in the outer
 Galaxy, where  the gas is supported by pressure against the net gravitational
field of the disc and the halo. It is then natural to ask what would be
the dynamical consequences of such a progressively prolate halo. This is the motivation for the present paper.

N-body simulations of galaxies show that the dark matter halo is triaxial. Along the Cartesian cuts (X-Y, or X-Z), the halos show a range of value from oblate to prolate with a preference for a prolate shape (Bailin \& Steinmetz 2005,
Bett et al. 2007, Vera-Ciro et al. 2011). Further the axis ratio varies with radius.
Indeed, there is no simple physical reason why the shape should be constant with radius. Thus a radially
variable axis ratio for an ellipsoidal mass distribution is not just of academic interest but of practical use 
for a study of dark matter halos, and 
in other astrophysical systems as well.

To study the dynamics of a prolate halo, we need to first calculate its potential and other physical quantities.
Mathematically this reduces to the problem of inversion of Poisson equation for a given density distribution.
 This is a rich and an old subject, which had attracted the attention of mighty mathematicians like MacLaurin, Lagrange, Jacobi, Gauss
to name a few (e.g. Rohlfs 1977), also see Chandraskekhar (1969) for details. Yet, interestingly, the 
calculation for a prolate case with  varying eccentricity
 has to our knowledge not been studied in the literature so far.

Motivated by this, in this paper we study prolate ellipsoidal mass distributions with an eccentricity that varies with radius. The ellipsoids are taken to be concentric and coaxial or coplanar.
We show that the constancy of potential within a shell as in homoeoids is  a good approximation 
even when the shell is bounded by dissimilar ellipsoids, 
for small variation in the eccentricity as seen in realistic systems.
This is an important general result from this paper.

 We calculate the potential for a general 
density distribution by adding the shell-by-shell contributions from isodensity surfaces from shells internal and external to a point, and from this obtain the radial and planar force equations and frequencies. 
Next, these results are applied to the specific case of pseudo-isothermal density distribution for the dark matter halo, and an exponential disc of finite thickness,
 to study the longevity of the Galactic warp
 for a disc in a prolate halo.  We show that the prolate halo does not result in a long-lived warp, contrary to
the literature where a prolate halo (Ideta et al. 2000) or a halo with a radially variable shape (Binney \& Tremaine 1987)
have been suggested to explain long-lived warps.

The calculation of the potential and related physical quantities are given in Section 2. In Section 3 this is applied to toy mass models so as to study the properties of  a prolate spheroidal system. Section 4 we apply this to the Milky Way and study the effect of the prolate halo on the warp and show that the halo does not yield long-lived warps. 
Section 5 summarizes our conclusions. 

The reader who is mainly interested in the application of a
prolate halo to the study of warps, may wish to skip the technical
material involving potential theory in Sections 2 and 3, and go
directly to Section 4.

\section{Potential of a Prolate Ellipsoid with Varying Eccentricity}

The dark matter halo is usually modeled  as a spherical or oblate distribution of 
constant ratio, for simplicity (e.g.,  Sackett \& Sparke 1990). In a rare
case, an oblate mass distribution with radially varying axis ratio has been studied
(Ryden 1990) in the quadrupolar limit, and extended for the prolate case by Buote \& Canizares (1996) in the
same limit.  The general case of a prolate spheroidal mass distribution, in particular with a
varying axis ratio, has not been studied in the literature so far. The potential for a general
triaxial case (with constant axis ratios) is stated in Binney \& Tremaine (1987) without derivation, and has been derived in Chandrasekhar (1969), and  by de Zeeuw \& Pfenniger (1988). From this
one can derive the potential for a prolate spheroid of constant axis ratio, as for example 
was done by Das \& Jog (1995) for modeling bars in a study of heating due bars.

In this paper, we give the analytical calculation of the potential and 
other relevant physical quantities of a body for which the isodensity surfaces are prolate in shape with radially varying eccentricity.  In this way of formulation of the problem, the case of a constant axis ratio comes out naturally
as a simple case of the general calculation, as shown later in this section.
First the potential due to a prolate spheroidal surface is calculated, following the approach
developed for an oblate spheroidal surface by Binney $\&$ Tremaine (1987). 
Then by summing over the contribution by the various shells, the net potential is obtained.

\subsection{Potential of a Prolate Spheroidal Surface}

To calculate the potential of a prolate spheroidal surface, it is appropriate to use the prolate spheroidal coordinate system $(u,v, \phi)$. However, the galactic disc is usually represented in cylindrical coordinates.
The prolate spheroidal coordinates $(u,v)$ and the cylindrical coordinates $(R,z)$ are related by: 
\begin{eqnarray}
 R&=&\Delta\,sinhu\,sinv
\nonumber\\
 z&=&\Delta\,coshu\,cosv
\label{eq:Rz}
\end{eqnarray}
where $\Delta$ is a constant (see e.g., Arfken 1970). \\ 

\noindent The Laplacian of the potential can be expressed in prolate spheroidal coordinates as:

\begin{eqnarray}
 \nabla^2{\Phi}=\frac{1}{h_{u}h_{v}h_{\phi}}\left [\frac{\partial}{\partial u}\left(\frac{h_vh_{\phi}}{h_u}\frac{\partial{\Phi}}{\partial u}\right)+ \frac{\partial}{\partial v}\left(\frac{h_uh_{\phi}}{h_v}\frac{\partial{\Phi}}{\partial v}\right)\right.
\nonumber\\
\left. +\frac{\partial}{\partial {\phi}}\left(\frac{h_vh_u}{h_{\phi}}\frac{\partial{\Phi}}{\partial \phi}\right) \right ] 
\end{eqnarray}

\noindent where the scale factors are:
\begin{eqnarray}
 h_u&=&\Delta (sinh^2u+sin^2v)^{1/2}
\nonumber\\
 h_v&=&\Delta (sinh^2u+sin^2v)^{1/2}
\nonumber\\
 h_{\phi}&=&\Delta (coshu+sinv)
\end{eqnarray}
Here the density is a function of $u$ only, so we expect the potential $\Phi(u)$ to be a function of 
only the coordinate $u$. For potentials of this class, after substituting
the scale factors in the eq. (2), $\nabla^2{\Phi}=0$
reduces to:
\begin{equation}
 \frac{d}{du}\left(sinhu\,\frac{d\Phi}{du}\right)=0
\end{equation}
Hence either 
\begin{equation}
 \Phi=\Phi_0 
\end{equation}
which is a constant, or 
\begin{equation}
\left(d\Phi/du\right)=A\,csch \: u
\end{equation}
 where $A$ is a constant. On integrating this equation we find, 

\begin{equation}
 \Phi=-A/2 \left[log\left(\frac{1+sech\,u}{1-sech\,u}\right)+log\psi_0\right]
\label{eq:phi}
\end{equation}
For large u, $sech\,u\,\rightarrow\,0$, and $sech u = \Delta /r$  so a potential of the form (eq. 7) varies as: 
\begin{equation}
 \Phi\simeq\,(-A/ 2) \left(2\,sech\,u + log\psi_0\right)\,\simeq\,-A\left(\frac{\Delta}{r}-log\psi_0\right)
\label{eq:phi2}
\end{equation}
where $r$ is the usual spherical radius. 

Hence, if we set $log \psi_0 =0$, and  $A=GM/\left(\Delta\right)$, the potential given by  eq. (8) tends to zero at infinity like the gravitational potential of mass $M$, as expected, here$M$ is the mass of the surface.
Thus we are led to consider,
\begin{equation}
\Phi =
\left\{
\begin{array}{ll}
 -\frac{GM}{2\ \Delta}\times\,log\left(\frac{1+sech\,u_0}{1-sech\,u_0}\right) & \mbox{when $u<u_0$} \\
\nonumber\\
 -\frac{GM}{2\, \Delta}\times\,log\left(\frac{1+sech\,u}{1-sech\,u}\right) & \mbox{when $u\geq u_0$}
\label{eq:totphi1}
\end{array}
\right.
\end{equation}

This potential is discontinuous on the spheroid $u=u_0$. Hence it is the gravitational potential of a shell of material on the surface 
$u=u_0$. This shell has  principal semi-major and semi-minor axes of lengths $c_0 \equiv\,\Delta\,coshu_0$ and $a_0\,\equiv\,\Delta\,sinhu_0$ respectively. Hence  eccentricity of the shell, $e=sech\,u_0$, thus the constant 
$\Delta = c_0  e$. We may rewrite eq. (9) as:
\begin{equation}
\Phi =
\left\{
\begin{array}{ll}
-\frac{GM}{2\,c_0\,e}\times\,log\left(\frac{1+e}{1-e}\right) & \mbox{when $u<u_0$},\\
\nonumber\\
-\frac{GM}{2\,c_0\,e}\times\,log\left(\frac{1+sech\,u}{1-sech\,u}\right) & \mbox{when $u\geq u_0$}
\end{array}
\right.
\label{eq:totphi2}
\end{equation}
\noindent On applying Gauss's theorem to the potential across this surface, 
its surface density (following the approach in Binney \& Tremaine 1987, see eq. 2-68)
is derived,
to be:
\begin{equation}
 \Sigma=\frac{\hat{n}_u\,.\,\nabla\Phi}{4\pi G}
\end{equation}
\noindent where, $\hat{n}_u$ is a unit vector normal to the surface, and $\nabla\Phi$ is the gradient of the potential:
\begin{equation}
\nabla\Phi=\frac{1}{\Delta ({sinh^2u+sin^2v})^{1/2}}\: \frac{d\Phi}{du}|_{u=u_0}
\end{equation}
\noindent Hence,
\begin{equation}
 \Sigma=\frac{M}{4\pi a^2 q ({1-e^2cos^2v})^{1/2}} 
\end{equation}

\subsection{Potential of Prolate Mass Distribution}
The potential due to a prolate spheroidal surface is obtained above (eq.[10]). A prolate spheroidal mass distribution can be considered 
as the combination of such surfaces, with the condition that the isodensity surfaces should not intersect each other. 
This is so that the function is not unphysical or double-valued.
Then 
we can calculate the potential of such distributions using eq. (10).

\noindent Define a quantity:
\begin{equation}
 \sigma\left(a\right)=\frac{a}{q}\,\frac{dq}{da}
\label{sigma}
\end{equation}
\noindent where $a$ is the semi-minor axis. The constraint that the shells do not cross is equivalent to stating that the semi-major axis of the outer shell must not be shorter than the 
semi-major axis of the inner shell, this requires that
$\sigma\left(a\right)\geq\,-1$.\\

Consider a thin shell of matter whose inner surface is a prolate spheroid with a semi-minor axis $a$ and axis ratio $q$, and outer 
surface is a prolate spheroid with a semi-minor axis $a+\delta a$ and an axis ratio $q+\delta q$. The mass of this prolate spheroidal shell, when $ \delta a << a $, can be shown to be:

\begin{equation}
 \delta M=4\,\pi \,a^2\,q\,\rho\left(a\right)\left[1+\frac{\sigma\left(a\right)}{3}\right]\,\delta a
\label{eq:mass}
\end{equation}

\noindent 
We next calculate $\bar\Sigma$, the surface density of  a prolate spheroidal shell with dissimilar surfaces
following the procedure in Binney \& Tremaine (1987, see their eq. 2-71) as:

\begin{equation}
 \bar{\Sigma}=\frac{\rho(a) \delta a}{|\nabla a|}
\end{equation}

\begin{equation}
\bar{\Sigma}=\frac{\rho(a) \delta a}{\left(R^2+\frac{z^2}{q^4}\right)^{1/2}} \left(a+\frac{z^2}{q^3}\frac{dq}{da}\right)
\end{equation}

\noindent where $R$ and $z$ are the cylindrical co-ordinates. On using eq. (1), and the discussion following eq. (9), these
reduce to:
\begin{eqnarray}
R=a\,sinv\\
 z=aq\,cosv
\end{eqnarray}
\noindent Hence, the surface density of the shell is given as:
\begin{equation}
\bar{\Sigma}=\frac{\rho\delta a}{a \left(1-e^2cos^2v\right)^{1/2} }\left[1+\sigma cos^2v\right]
\end{equation}
\noindent Using eq. (15) for the mass of the shell, eq. (20) reduces to:
\begin{equation}
 \bar{\Sigma}=\frac{\delta M}{4\pi a^2 q \left(1-e^2cos^2v\right)^{1/2} }\left[\frac{1+\sigma \,cos^2v}{1+ (\sigma/3)}\right]
\end{equation}

\noindent Note that for similar surfaces, $\sigma = 0$ (eq. [14]), hence the surface density of a shell within two similar surfaces obtained by taking a projection of the mass on the inner surface of the shell (eqs.[16] and [21]) is identically equal to that obtained using the Gauss's theorem approach (eq.[13]). Hence the potential inside a shell bounded by similar surfaces can be taken to be constant (Newton's theorem), or there is no force at an internal point due to such a shell. This is the standard homoeoidal theory of ellipsoids (e.g., Chandrasekhar 1969).

It is not clear a priori that the results from the homoeoidal potential theory are applicable for a shell bounded by  dissimilar surfaces considered here. In fact, in this case,
 the surface density obtained by taking the projection (eq. [21]) differs from that obtained earlier by the Gauss's theorem approach
(eq.[13]) by a dimensionless scale factor $S$, defined as:

\begin{equation}
S = \left[\frac{1+\sigma \,cos^2v}{1+ (\sigma/3)}\right]
\end{equation}

This scale factor
 denotes the deviation from the case of constant 
internal potential. We show in Appendix A that the values of $S$ are small 
for the small variation in eccentricity with radius considered here. Thus, interestingly the homoeoidal potential theory though not
rigorously valid, is still a good approximation.
Hence the potential within a shell  ($u < u_0$, see eq.[10]) can be assumed to be constant
even for the case of a shell bounded by dissimilar spheroids. In analogy with a homoeoid which is a shell bounded by two similar spheroids, we define a heteroid to be a shell bounded by two dissimilar spheroids. In this paper we study the gravitational potential and force due to a heteroid.

Hence, for a field point lying inside the spheroid of semi-minor axis $a$ (with a prolate co-ordinate $u_a$), the contribution of potential at that point due to this shell (from eqs. [9] and [15]) is given as: 
\begin{equation}
 \delta \Phi_{int}=-2\pi G\frac{a}{e}\left[1+\frac{\sigma\left(a\right)}{3}\right]log\left(\frac{1+sechu_a}{1-sechu_a}\right)\rho\left(a\right)\delta a
\label{eq:phi_int}
\end{equation}
From eq.(23) onwards, $e \equiv  sech u_a$.

Similarly if the point lies outside the shell, then
\begin{equation}
 \delta \Phi_{ext}=-2\pi G\frac{a}{e}\left[1+\frac{\sigma\left(a\right)}{3}\right]log\left(\frac{1+sech u_a''}
{1-sech u_a''}\right)\rho\left(a\right)\delta a 
\label{eq:phi_ext}
\end{equation}
\noindent where $u_a''$ is the label of the spheroid that is confocal with the spheroid with semi minor axis $a$ and lies outside of it.
 
The potential of the entire body is the sum of contributions from all the homoeoids which make up the body. The contribution from the outer shells at a point $(R,z)$ is: 
\begin{eqnarray}
\sum_{a>a'}\delta \Phi_{int}=-2\pi G\int_{a'}^\infty\frac{a}{e}\left[1+\frac{\sigma(a)}{3}\right]\;\;\;\;\;\;\;\;\;\;\;\;\;\;\;\;\;\;\;
\nonumber\\
\times \,log\left(\frac{1+sechu_a}{1-sechu_a}\right)\rho(a)\,da 
\end{eqnarray}

The contribution from the inner shells to the potential at the point $(R,z)$ is:
\begin{eqnarray}
\sum_{a<a'}\delta \Phi_{ext}=-2\pi G\int_{0}^{a'}\frac{a}{e}
\left[1+\frac{\sigma(a)}{3}\right]\;\;\;\;\;\;\;\;\;\;\;\;\;\;\;\;\;\;\;
\nonumber\\
\times\,log\left(\frac{1+sechu_a''}{1-sechu_a''}\right)\rho(a)\,da
\end{eqnarray}
where $a'$  is given by:
\begin{equation}
{a'}^2=R^2+\frac{z^2}{q^2\left(a'\right)}
\end{equation}
The total potential, $\Phi$, is given by the summation of eqs. (15) and (16) as:
\begin{equation}
\Phi=\sum_{a>a'}\delta \Phi_{int}\,+\,\sum_{a<a'}\delta \Phi_{ext}\,
\label{eq:totphi}
\end{equation}

In the limit of a constant axis ratio, $\sigma = 0$, hence the above expression simplifies to the result for the prolate spheroidal mass distribution of a constant shape.

\subsection {Related Physical Quantities}
Using the above expression for the potential, we next obtain other relevant physical quantities such as the force components along $R$ and $z$, and the angular and vertical frequencies at the mid-plane for a particle in this potential. These are 
applied  to the particular problem of warps in Section 4.
\subsubsection {Angular Frequency, $\Omega_h$}
By taking the gradient of eq.(18), we can calculate the force in horizontal and vertical directions. 
Then force in the 
radial direction is:
\begin{eqnarray}
 F_R\left(R,z\right)=2\pi G\frac{\partial}{\partial R}\int_{0}^{a'}\frac{a}{e}\left[1+\frac{\sigma(a)}{3}\right]\;\;\;\;\;\;\;\;\;\;\;\;\;\;\;\;\;\;\;
\nonumber\\
\times\,log\left(\frac{1+sechu_a''}{1-sechu_a''}\right)\rho(a)\,da
\label{eq:fr}
\end{eqnarray}
The contribution of force from the outer shells is zero since the potential is constant inside such shells. To take the gradient in the cylindrical coordinate system, we rewrite $sechu_a''$ in terms of $R$ and $z$, by using the relation,
\begin{equation}
\left({aqe}\right)^2=\frac{R^2{sech^2u_a''}}{1-sech^2u_a''}+z^2{sech^2u_a''}
\label{eq:sech}
\end{equation}
At the galactic mid-plane ($z=0$), eq. (30) reduces to:
\begin{equation}
 sechu_a''=\frac{a\,q\,e}{(R^2+a^2\left(q^2-1\right))^{1/2}}
\label{eq:sech2}
\end{equation}
Substituting eq. (31) in eq. (29), and taking the derivative in the radial direction, gives:
\begin{equation}
F_R\left(R\right)=\frac{4\pi G}{R}\int_{0}^{R}a^2q\left[1+\frac{\sigma(a)}{3}\right]\frac{\rho (a)}{[R^2+a^2\left(q^2-1\right)]^{1/2}}da
\label{eq:fr0}
\end{equation}
Notice that  at the galactic plane, $a' = R$.\\
Then the square of the rotation velocity ($V^2_c=F_R R $) is: 
\begin{equation}
V^2_c\left(R\right)=4\pi G\int_{0}^{R}a^2q\left[1+\frac{\sigma(a)}{3}\right]\frac{\rho (a)}{[R^2+a^2\left(q^2-1\right)]^{1/2}}da
\label{eq:vcp}
\end{equation}
\noindent The square of the angular frequency, ($\Omega^2={F_R}/{R})$, is:
\begin{equation}
\Omega_h^2\left(R\right)=\frac{4\pi G}{R^2}\int_{0}^{R}a^2q\left[1+\frac{\sigma(a)}{3}\right]\frac{\rho (a)}{[R^2+a^2\left(q^2-1\right)]^{1/2}}da
\label{eq:omp}
\end{equation}
where the subscript $h$ denotes the halo.
\subsubsection{Vertical Frequency, $\nu_h$}
\noindent The force in the vertical direction is obtained by taking a gradient of eq. (18) along $z$:
\begin{eqnarray}
 F_z\left(R,z\right)=2\pi G\frac{\partial}{\partial z}\int_{0}^{a'}\frac{a}{e}\left[1+\frac{\sigma(a)}{3}\right]\;\;\;\;\;\;\;\;\;\;\;\;\;\;\;\;\;\;\;
\nonumber\\
\times\,log\left(\frac{1+sechu_a''}{1-sechu_a''}\right)\rho(a)da
\end{eqnarray}
On simplifying,
\begin{equation}
 \frac{\partial}{\partial z} \log \frac{1+sechu_a''}{1-sechu_a''}=2\,\frac{\frac{\partial sechu_a''}{\partial z}}{1-sech^2u_a''}
\end{equation}
On taking the derivative of eq. (30) in the vertical direction, we get:
\begin{equation}
 \frac{\partial sechu_a''}{\partial z} =\,\frac{z\,\left(sechu_a''-sech^3u_a''\right)}{2z^2\,sech^2u_a''-\left(z^2+R^2+a^2\left(q^2-1\right)\right)}
\label{eq:dsech}
\end{equation} 
Substituting  eqs. (36) and (37) into eq. (35), and 
taking the limit $z \rightarrow 0$ gives:
\begin{equation}
 F_z(R,z)=4\pi Gz\int_{0}^{R}a^2q\left[1+\frac{\sigma(a)}{3}\right]\frac{\rho (a)}{(R^2+a^2\left(q^2-1\right))
^{3/2}}da
\label{eq:fz0}
\end{equation}
Then the square of the vertical frequency, $\nu^2={F_z}/{z}$ is: 
\begin{equation}
 \nu_h^2\left(R\right)=4\pi G\int_{0}^{R}a^2q\left[1+\frac{\sigma(a)}{3}\right]\frac{\rho (a)}{\left(R^2+a^2\left(q^2-1\right)\right)^{3/2}}da
\label{eq:nup}
\end{equation}
For the sake of completeness, in Appendix B a similar calculation is given for an oblate spheroid with varying eccentricity.

\section {Application for various model mass distributions}
In Section 2, we derived the expressions of potential, forces, angular frequency
and vertical frequency of prolate shaped mass distributions.
In order to get an idea of how 
these quantities vary with radius, we now apply those equations to some simple
toy models of mass distributions. We will consider that the density distribution varies isothermally as:
\begin{equation}
 \rho(a)=\frac{\rho_0}{1+\frac{a^2}{a_c^2}}
\label{eq:rho_a}
\end{equation}
where $\rho_o $ and $a_c$ are the core density and core radius respectively (Binney \& Tremaine 1987).\\
The isodensity surfaces are taken to be aligned prolate spheroids whose axis ratio has
the value:
\begin{equation}
 q(a')=\beta(1+\frac{a}{a_c})^\alpha
\label{eq:q}
\end{equation}
where $a/a_c$ is the dimensionless semi-minor axis and $\alpha\,\&\,\beta$ are constants. Within the constraint
$\sigma(a/a_c)\geq -1$ (eq.[14]), we use the values of $\alpha=-0.1,0 $ and $0.1$ and value of $\beta=1.5$. Fig. 1 shows the isodensity contours
of these mass distributions. For $\alpha=0$ we get isodensity contours of constant axis ratio $(q=1.5)$ for all radii (Fig. 1a). For $\alpha=0.1$, the spheroids become more prolate at larger radii (from $q=1.5$ to $1.9$) (Fig. 1b), and for $\alpha =-0.1$ 
the isodensity surfaces become less prolate  (from $q=1.5$ to $1.2$) (Fig. 1c).

\begin{figure}
 \centering    
\includegraphics [height=4.0cm,width=4.0cm]{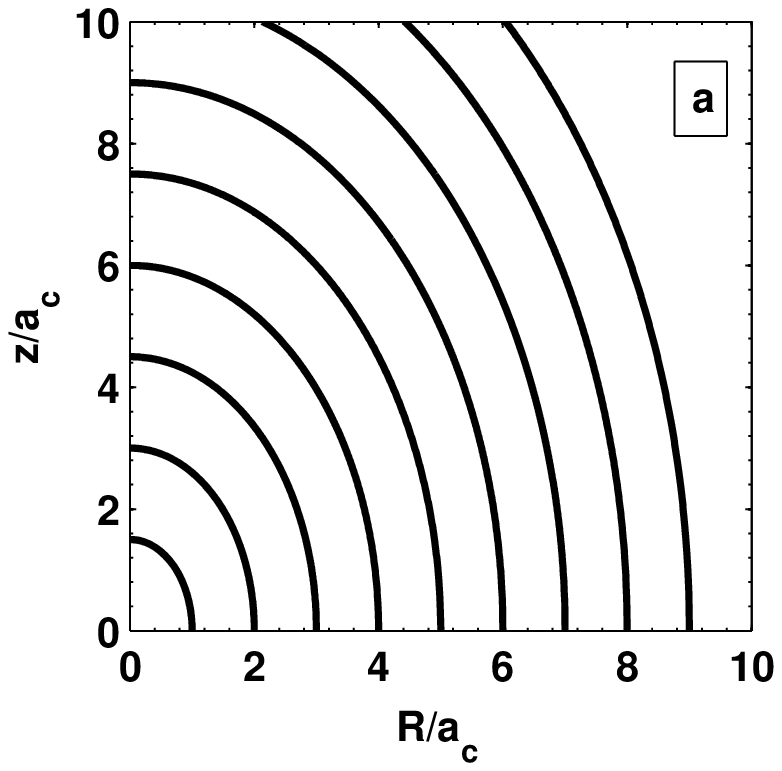}
\includegraphics [height=4.0cm,width=4.0cm]{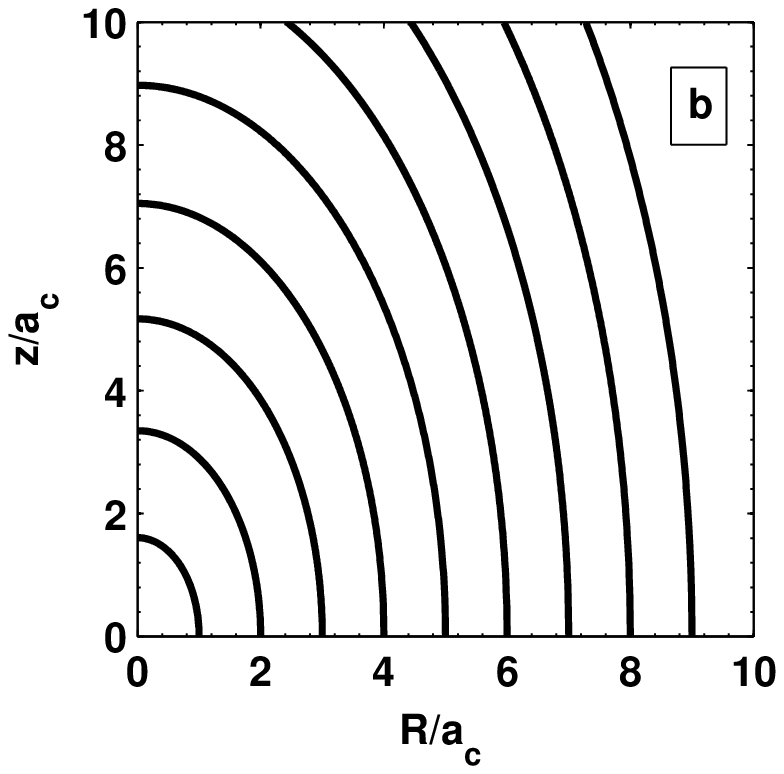}
\hspace{3mm}
\includegraphics[height=4.0cm,width=4.0
cm]{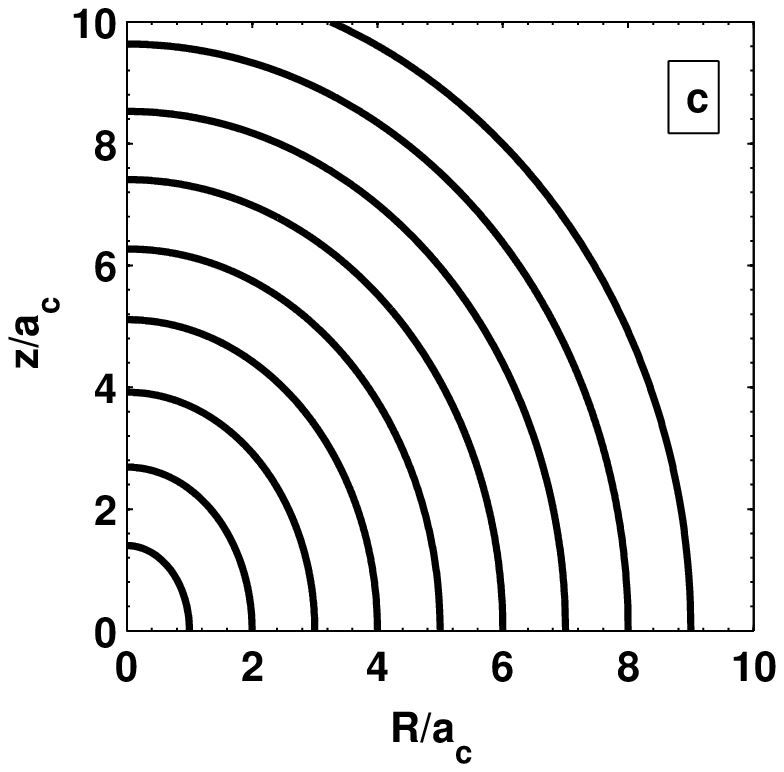}
\caption{
        Isodensity contours in a plane containing the polar axis for a prolate spheroidal mass distribution  with an axis ratio that is (a) constant, (b) increasing, and (c) decreasing with radius.
     }
\end{figure}

We next obtain the rotation curves for the three cases considered.
For a density profile with a constant axis ratio,  the expression for the square of the rotation velocity ${V_c}^2$ can be obtained analytically by integrating eq. (33) as:
\begin{equation}
V_c^2=\frac{4{\pi}G\rho_0a_c^2}{e}\left[\frac{1}{2}log\left(\frac{1+e}{1-e}\right)-
\left(\frac{c}{b}\right)^{1/2}\arctan{ \: e\left(\frac{b}{c}\right)^{1/2}}\right]
\end{equation}
At large radii, this is independent of radius, and in the limit of 
$R\rightarrow\infty,\ $ the constant terminal velocity square is given by:
\begin{equation}
 V_H^2=\frac{2{\pi}G\rho_0a_c^2q}{({q^2-1})^{1/2}}\,log\left(\frac{1+e}{1-e}\right)
\label{eq:vH}
\end{equation}
where $c$ and $b$ are defined as: 
\begin{eqnarray}
c &=& a_c^2(q^2-1)
\nonumber\\
b &=& R^2-a_c^2(q^2-1)
\label{eq:bc}
\end{eqnarray}
For the case of varying axis ratio, rotation velocity is calculated numerically. Fig. 2 shows the rotation velocity curve 
for $\alpha=0, +0.1$ and $-0.1$.
The \% differences in resulting velocities at large radii for the various values of $\alpha$ are not very large.
\begin{figure}
\centering
\includegraphics[height=4.8cm,width=6.5cm]{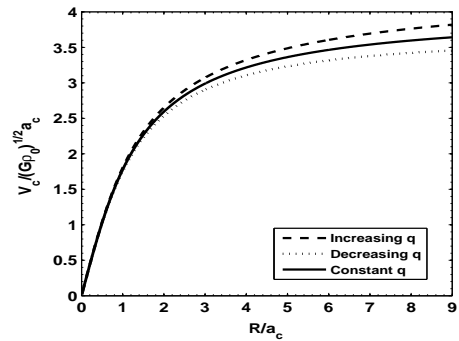}
\caption{The rotation velocity for a pseudo-isothermal prolate-shaped halo density distribution with an axis ratio that is: constant ($q=1.5$), increasing ($q=1.5$ to $1.9$) and decreasing ($q=1.5$ to $1.2$) with radius.}
\end{figure}
In Section 4, we will apply these results to study the pattern speed and winding of the Galactic warp. Here we  calculate  the pattern speed of warp due to dark matter halo alone. 
In the kinematical
picture of warps, the pattern speed for the slow mode is given by $\Omega - \nu$ where $\Omega$ and $\nu$ represent the
angular speed and the vertical frequency respectively (e.g., Binney \& Tremaine 1987). This is obtained by treating the warp as a small perturbation and keeping only the slow mode since the fast mode will wind up faster.
The pattern speed 
 of a galactic warp due to dark matter halo alone, ($(\Omega_p)_h$), is defined by:
\begin{equation}
 (\Omega_p)_h=\Omega_h-\nu_h
\label{eq:om}
\end{equation}
where the subscript $h$ represents the halo contribution. 
 Using the density profile and the axis ratio of the dark matter halo, as a function of the semi-minor axis from eqs. (40) and (41), and substituting these in eqs. (34) and (39),
gives the square of the angular and vertical frequencies. 
For the case of a constant axis ratio,  
${\Omega_h}^2 = {V_c}^2 / R^2$ where ${V_c}^2$ is as given by eq. (42), and the vertical frequency square is given by
integrating eq. (39) to be:
\begin{equation}
\nu_h^2=\frac{4{\pi}G\rho_0a_c^2}{be}\left[ e-{\left(\frac{c}{b}\right)^{1/2}}\arctan { e \left({\frac{b
}{c}}\right)^{1/2}}\right]
\end{equation}
where $b$ and $c$ are defined in eq. (44).\\

For density profiles with a varying axis ratio, $\Omega_h$ and $\nu_h$ are calculated numerically. 
Fig.~3 shows the pattern speed curves of different distributions: the pattern speed decreases at different rates with radius and approaches zero at large radii. For each axis ratio considered, the mass distribution near the mid-plane is different  which mainly determines the rate of fall of the pattern speed curves.
The important result to note from Fig. 3 is that for a prolate halo the pattern speed is $>0$, that is, 
$\Omega_h > \nu_h$. In contrast, for a flattened disc or an oblate halo, the reverse is true (e.g, Binney \& Tremaine 1987).  
The net value of the pattern speed is determined by taking account of the
 contribution from the galactic disc also, as will be discussed in Section 4.  
\begin{figure}
\begin{center}
\includegraphics[height=4.8cm,width=6.5cm]{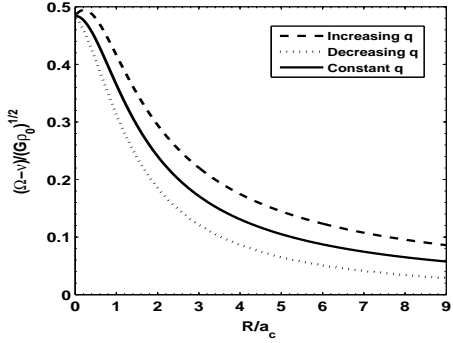}
\end{center}
\caption[]{The pattern speed for the halo-alone case, for a
prolate-shaped halo density distribution with an axis ratio that is: constant ($q=1.5$), increasing ($q=1.5$ to $1.9$) and decreasing ($q=1.5$ to $1.2$) with radius.}

\end{figure}

\section {Application to warp in the Galaxy}

Here we apply the results obtained in Section 2 to the particular case of the Milky Way to study the  pattern speed and winding in the Galactic warp. The aim is to see what effect a progressively prolate halo has on the winding time of a warp.

In the past it has been suggested that a prolate halo (Ideta et al. 2000), or an oblate halo with an eccentricity that changes with radius (Binney \& Tremaine 1987) can help in making warps long-lived.
The 
severity of the winding problem depends on the differential precession rate or on how rapidly the pattern speed changes with radius. In the kinematical
picture of warps, the pattern speed for the slow mode is given by $\Omega - \nu$ 
(Section 3). We need to calculate the contribution to this both from the disc and the halo. In a disc, the pattern speed is negative since $\nu > \Omega$ for a gravitating disc, while for a spherical halo the two are equal, and for a prolate halo the reverse is true, namely, $\nu < \Omega$.
Thus it can be seen that adding a prolate halo may help sustain a warp. 
Qualitatively this was the motivation for trying a prolate halo in the present paper.

\subsection{Vertical and angular frequencies due to Galactic disc}

For most dynamical studies, the Galactic disc is considered to be infinitesimally thin for simplicity.
However, the vertical force is then discontinuous at the galactic plane, hence it is not possible to approximate small displacements in the vertical direction perturbatively as a simple harmonic motion.
 Hence we have to use a  three dimensional density distribution.
Here we use the simple double exponential density distribution for the Galactic disc which is given by:
\begin{equation}
\rho\left(R,z\right)=\rho\left(0,0\right)\,exp\left({-\frac{R}{R_d}}\right)\,exp\left({-\frac{z}{z_d}}\right)
\end{equation}
where $\rho(0,0)$ is the central density; $R_d$ and $z_d$ are the radial and vertical disc scale lengths respectively.

The corresponding potential at a point $(R,z)$ is given by Kuijken \& Gilmore (1989), also see Sackett \& Sparke (1990), and Cuddeford (1993):
\begin{eqnarray}
\Phi_d\left(R,z\right)=-4{\pi}G\rho\left(0,0\right) R_d^2 z_d \;\;\;\;\;\;\;\;\;\;\;\;\;\;\;\;\;\;\;\;\;\;\;\;\;\;\;\;\;\;\;\;\;\;\;\; \nonumber \\
 \times \int_0^{\infty}\frac{J_0\left(kR\right)}
{\left(1+\left(kR_d\right)^2\right)^{3/2}} 
 \frac {[k z_d \, exp^{{- \vert z \vert}/ z_d} - exp^{- k \vert z \vert]}}{k^2 z_d^2 - 1}dk 
 \end{eqnarray}
\noindent where $J_0(x)$ is the first order Bessel function of order zero.

The radial force $F_R(R,z)$ is given by $- \partial \Phi_d / \partial R$ ,and then taking the limit as $z \rightarrow 0$,
we get its value at mid-plane to be (also see, Kuijken \& Gilmore 1989):
\begin{eqnarray}
 F_R(R,0)=4{\pi}G\rho(0,0)R_d^2z_d\;\;\;\;\;\;\;\;\;\;\;\;\;\;\;\;\;\;\;\;\;\;\;\;\;\;\;\;\;\;\;\;\;\;\;\;\;\;
\nonumber\\
\times\,\int_0^{\infty}\frac{kJ_1\left(kR\right)}{\left(1+kz_d\right)\left(1+\left(kR_d\right)^2\right)^{3/2}}\,dk
\end{eqnarray}

\noindent Similarly, for the force along the vertical direction at the mid-plane, we obtain $F_z (R,z) = - \partial \Phi_d / \partial z$ and take its limit at z=0, to get:

\begin{eqnarray}
 F_z(R, z\rightarrow 0)=4{\pi}G\rho\left(0,0\right)R_d^2z\;\;\;\;\;\;\;\;\;\;\;\;\;\;\;\;\;\;\;\;\;\;\;\;\;\;\;\;\;\;\;\;\;\;\;\;
\nonumber\\
\times\int_0^{\infty}\frac{kJ_0\left(kR\right)}{\left(1+kz_d\right)\left(1+\left(kR_d\right)^2\right)^{3/2}}dk
\end{eqnarray}

\noindent From these equations of force (eqs. [49] and [50]), the angular and vertical frequencies are given by the expressions:
\begin{equation}
{\Omega_d} ^2  = {F_R (R,0)}/{R} 
\end{equation}
and,
\begin{equation}
{\nu_d} ^2  =  {F_z (R, z \rightarrow 0)}/{z} 
\end{equation}

\subsection {Disc and halo parameters for the Galaxy}
For the Galaxy, the radial disc scale length $R_d=3.2\,kpc$ (Mera et al. 1998) and $z_d = 0.35\,kpc$ (Binney $\&$ Tremaine 1987) are used for calculating the force components and the frequencies given above.
The mid-plane volume density of the stellar disc at the solar neighborhood is observed to be  $0.07 M_{\odot}\,pc^{-3}$ (Binney $\&$ Tremaine 1987). Substituting the values of $R_d$, $z_d$ and $\rho$ at solar neighbourhood in eq. (47), we get the value of the central density, $\rho(0,0) \approx 0.99 \,M_{\odot}\,pc^{-3}$.

Three types of halo distributions are considered:
\\
(i) A pseudo-isothermal spherical distribution as in eq. (40), obtained in the Galactic mass model by Mera et al (1998).\\ where $\rho_0=0.035 \,M_{\odot}\,pc^{-3}$ and $a_c=5 kpc$.  \\
(ii) A prolate spheroidal distribution with an increasing axis ratio, that is obtained by modeling the steep  flaring of HI in the outer Galaxy (Banerjee $\&$ Jog 2011); where 
axis ratio, $q$ and the halo density $\rho (a)$ are given respectively by:
\begin{equation}
q = 1+\alpha_1\,(a-9)+\alpha_2\,(a-9)^2  \mbox{when $9 kpc \leq a\leq 24 kpc$}, \\
\end{equation}
\noindent and,
\begin{equation}
\rho(a)=\frac { {{\rho_0}/\left[{1+{(\frac{a}{a_c})}^2}\right]}}{q} 
\end{equation}
\noindent where $q=1$ when $a\leq9 kpc$, and the constants $\alpha_1=0.020$ and $\alpha_2=0.003$,
and  $\rho_0=0.035 \,M_{\odot}\,pc^{-3}$ and $a_c=5 kpc$.\\
(iii) A prolate shape with a constant axis ratio, $q$ =1.9 :
\noindent We change the shape of the dark matter halo from a sphere to a prolate spheroid  by keeping the semi-minor axis fixed such that the total mass inside does not change. 
Thus equating the mass inside both spheroids,
one can find out the value of the core radius $a_c$ for the prolate spheroid. The mass enclosed by the prolate spheroid of semi minor-axis $a$ is: 
\begin{equation}
 M(<a)=4{\pi}G\rho_0a_c^2q(a-a_c\arctan(\frac{a}{a_c}))
\label{eq:masp}
\end{equation}
Equating this with the mass inside a sphere (i.e. M(q=1)) of radius $a_c$, gives the 
value of $a_c$ as a function of $q$:
\begin{equation}
 \frac{a_c(q)}{a_c(1)}\arctan\frac{a_c(1)}{a_c(q)}=1-\frac{4-{\pi}}{4 (q^2-1)^{1/2}}\,\frac{1}{2}\,log\left(\frac{1+e}{1-e}\right)
\end{equation}
Equating the terminal velocities of the prolate spheroid (eq. [43]) and of the sphere gives the value of $q (a)$: 
\begin{equation}
\frac{\rho_0(q)}{\rho_0(1)}=\frac{a_c^2(1)}{a_c^2(q)}\,\frac{({1-q^2})^{1/2}}{q}\,\frac{1}{\frac{1}{2}\,log\left(\frac{1+e}{1-e}\right)}
\end{equation}
where $a_c(1)=5\,kpc$ and $\rho_0(1)=0.035\,M_{\odot}\,{pc}^{-3}$ for the spherical case ($q=1$) (Mera et al. 1998). 
  For $q = 1.9$, this gives $a_c = 5.98$ kpc and $\rho_0=0.016\,M_{\odot}pc^{-3}$. For another typical value, $q=1.5$, we get
$a_c = 5.57$ kpc and $\rho_0=0.021\,M_{\odot}pc^{-3}.$

\subsection{Pattern Speed for Warp}
The pattern speed for a galactic disc embedded in a dark matter halo is set by adding in quadrature the contributions from both, and is defined to be: 
\begin{equation}
 \Omega_p= [{(\Omega_h^2\,+\,\Omega_d^2)\,-\,(\nu_h^2\,+\,\nu_d^2)}]^{1/2}
\end{equation}
Using the values of the disc and the three cases of halo parameters as discussed above, the resulting pattern speed curves are shown in Fig. 4. The disc determines the main behaviour of the pattern speed, hence the latter is $< 0$.
Surprisingly, the presence of dark matter halo 
makes little difference to the pattern speed curve. This is in contradiction
to the expectation from the literature that halo would affect the pattern speed significantly (Section 4.5). 
\begin{figure}
\begin{center}
\includegraphics[height=4.8cm,width=6.5cm]{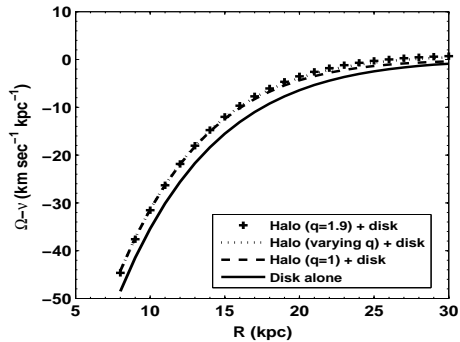}
\end{center}
\caption{The pattern speed of warp in the Galaxy  consisting of an exponential disc of finite height and dark matter halo of different shapes and
distributions. Clearly, the halo contribution to the 
pattern speed is negligible.}
\label{pattern1}
\end{figure}

\subsection{Winding Time}
The winding time 
 of a warp between the radii $R_1$ and $R_2$ is defined in terms of the pattern speeds at these two radii as:
$$ t_{winding} = \pi / [ \Omega_p (R_1) -\Omega_p (R_2) ] \eqno (59) $$
\noindent In order to quantify how rapidly warp gets wound up, we need to calculate the winding time using eq. (58).
If the Milky Way were to contain only the galactic disc, then the winding time of the warp that extends from $3R_d$ to $8R_d$ is obtained to be  $\sim 9.2\,\times \,10^7$ yr.
That is, the warp will be wound up within the age of the Galaxy ($\approx 10^{10}$ yr). 
If the galactic disc is embedded either in a pseudo-isothermal spherical halo or in a prolate spheroidal halo, the winding time is almost the same (to within a few percent). This shows that the dark matter halo with parameters as chosen here 
does not make any difference in the winding time. Hence, if the warp is considered as a bending wave in the galactic disc, the gravitational field of the
 dark matter halo does not solve the winding problem of the Galactic warp.

\subsection{Comparison with previous work}

In the past, it has been claimed (Binney $\&$ Tremaine 1987, Ideta et al. 2000) that the dark matter halo could increase the winding time of kinematic bending wave for a particular choice of halo parameters.
In these, the galactic disc was taken to be infinitesimally thin, with a surface density that falls exponentially with radius:
$$
\Sigma(R)=\Sigma_0 e^{-{R}/{R_d}} \eqno (60)$$
\noindent where $R_d$ is the disc scale length (the same as in eq. [47]) and 
$\Sigma_0$ is the central surface density.

 For such a disc, at radii much larger than the outer edge of the disc, the potential can be approximated as a sum of a monopole and a quadrupole term as shown by Binney \& Tremaine (1987) to be:
$$ \Phi_d=-\frac{GM_d}{r}\left[1+\frac{3R_d^2(R^2-2z^2)}{2r^4}\right] \eqno (61) $$
where $r$ is the radius in spherical coordinate system, and $M_d$ is the total mass of the disc ($M_d = 2 \pi \Sigma_0 
{R_d}^2$). 
The horizontal and vertical frequencies are given by:
$$\Omega_d^2=\frac{GM_d}{R^3}\left[1+\frac{9R_d^2}{2R^2}\right] ; \nu_d^2=\frac{GM_d}{R^3}\left[1+\frac{27R_d^2}{2R^2}\right] \eqno (62) $$

\noindent We caution that these are strictly valid only at large radii ($>>$ the size of the disc), whereas in the above references, these were
incorrectly applied to lower radii within the disc $\sim$ 10 kpc where the warp begins (see, Ideta et al. 2000, Fig. 8;
and Binney \& Tremaine 1987, Fig. 6-30). Both these references use a logarithmic potential for the halo while we have derived a rigorous expression for the halo potential.
Fig. 5 shows the results obtained using this approximation (dashed line), and that obtained in this paper (eqs. [51]-[52]) for a disc of finite thickness (solid line).
 Here we used $R_d$=3.2 kpc 
(see Section 4.1) and $\Sigma_0=640\,M_{\odot}\,pc^{-2}$ (Binney \& Tremaine 1987) for plotting the curves.
\begin{figure}
\includegraphics[height=4.8cm,width=6.5cm]{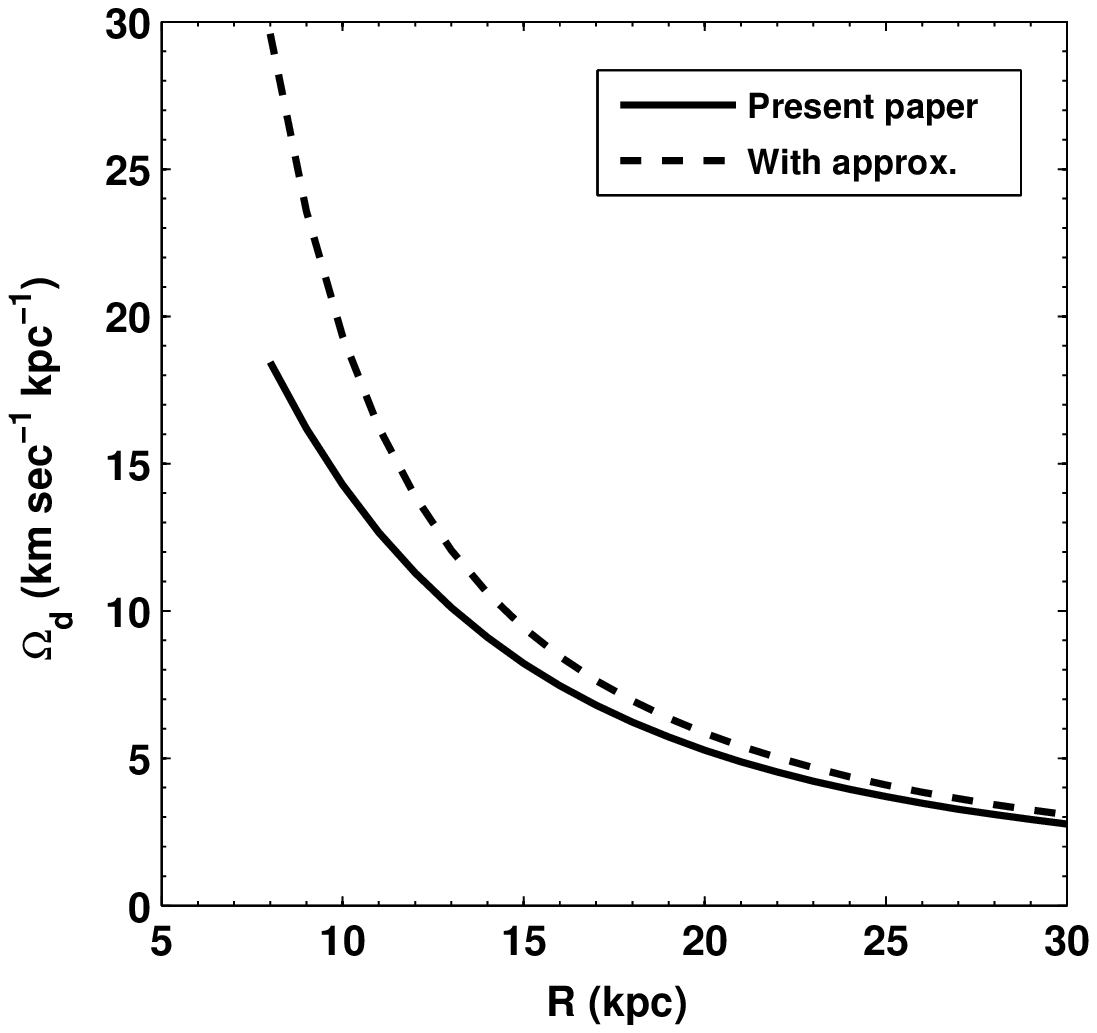}
\includegraphics[height=4.8cm,width=6.5cm]{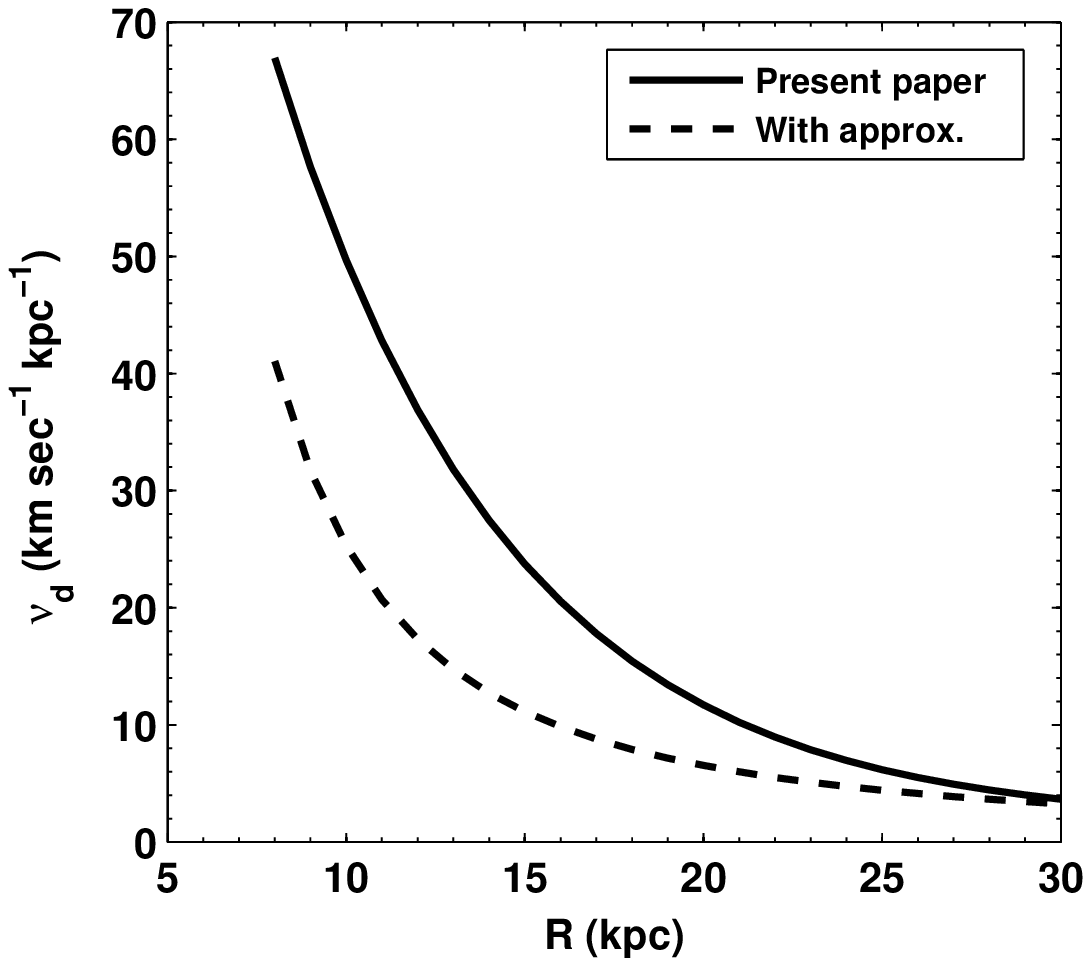}
    \caption{$\Omega_d$  and $\nu_d$ vs. R for a particle in the Galactic disc using the correct treatment as in this paper (solid line), and using the approximation of an infinitesimally thin disc and potential valid at large radii as in Binney \& Tremaine (1987) (dashed line).}
\label{pattern}
\end{figure}

In this approximation, $\nu$ and $\Omega$ have comparable values - we note that this result is incorrect, since it is well-known that within a self-gravitating disc $\nu >> \Omega$. Check that the correct treatment for a finite height disc 
 in the present paper does satisfy this criterion (see Fig. 4 which shows $\Omega - \nu < 0$), which results in a pattern speed that 
varies rapidly with radius and is dominated by the disc contribution (Fig. 4). 
\begin{figure}
\includegraphics[height=4.8cm,width=6.5cm]{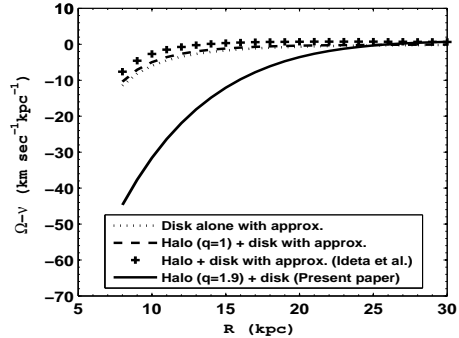}
\caption{The pattern speed of warp in the Galaxy consisting of a disc of finite thickness and a prolate-halo (q=1.9)
from the present paper (solid line), and that using the approximation of
 an infinitesimally thin disc and potential valid only at large radii (as in Binney \& Tremaine 1987, and Ideta et al. 2000) and different halo
shapes.}
\end{figure}

Figure 6 shows results for the pattern speed of warp at the galactic plane for a prolate halo of $q=1.9$ and the correct treatment for the disc as in this paper (the solid line), and it also shows the erroneous results obtained using the approximation for the disc (eq.[62]) either by itself, or with a spherical halo, and the logarithmic halo as in Ideta et al. (2000). The last three cases, in particular the one with the prolate halo, show an almost constant pattern speed throughout the range of radii of interest where the warp is seen.
This led
Ideta et al (2000) to the conclusion that a prolate halo results in a Galactic warp that
can be sustained for more than the Hubble time. To summarise, they arrived at this wrong conclusion due to the use of an approximate expression for the potential for an infinitesimally thin disc as applied to a low radial range where it is not valid.

The correct treatment for the disc in the present paper 
 shows that the disc contribution to the pattern speed is dominant (solid line, Fig. 6, also see Fig. 4), and that
 the prolate dark matter halo, indeed halo of any shape, has  very little effect 
on the pattern speed of the Galactic warp. 
 The normalized 
  differential force due to the heteroid shells considered is negligible ($< 1 \%$, see Appendix A), hence that too cannot affect the above result.
In retrospect, 
this is not surprising, since the previous work on the  origin of warps by non-axisymmetric response 
showed that the radius of onset of warps in a galactic disc depends weakly on whether or not an oblate halo is included or on its shape, as shown for a disc
with an exponential vertical distribution (Saha \& Jog 2006) or a general $z$ distribution (Pranav \& Jog 2010).

We caution that we have only considered a simple kinematical model of warp, as an application of the potential obtained for prolate ellipsoids of varying eccentricity. The latter was the main aim of this paper. A complete treatment of  the dynamics and origin of warps would be much more complex and could involve a  responsive and triaxial halo (e.g., Binney, Jiang \& Dutta 1998, Dubinski \& Chakrabarti 2009) and is beyond the scope of this paper.

\section {Conclusions}

In this paper, we have given the  analytical calculation for the potential 
of a prolate ellipsoidal mass distribution with radially varying eccentricity.
This is obtained by summing up the shell-by-shell contributions of isodensity surfaces. 
We define a heteroid
to be a shell that is bounded by two dissimilar surfaces.
An important result from this paper is that the potential within a heteroid is shown to be approximately constant
as long as the radial variation and the range of values of eccentricity spanned are small as in realistic systems.  The normalized differential
force across such a shell is non-zero but small  and does not affect the dynamical stability of such figures on a Hubble time.

 The motivation for this work was to see if the progressively prolate halo for the Galaxy  (as in Banerjee \& Jog 2011) could help sustain long-term warps.
We apply the above calculation to the case of Galactic warp, and on using the correct treatment for the disc, find that the disc mainly determines the warp pattern speed. Thus,  the dark matter halo has only a marginal effect on the pattern speed, and it does not lead to long-lived warps.

The results for the potential and force components obtained here are general  
and could be applied for a more
realistic treatment of other astrophysical systems such as prolate bars with a varying shape. 
We also give a similar exact calculation for the oblate case for the sake of completeness. This can be applied to oblate spheroidal systems with varying eccentricity.

\bigskip

\noindent  {\it Acknowledgements:} 

We thank the anonymous referee for raising the very important issue about the applicability of the homoeoid potential theory for the case of  spheroids of varying eccentricity studied in our paper.

\bigskip

\noindent {\bf References}

\bigskip

\noindent Arfken, G.B.  1970, Mathematical methods for physicists, New York: Academic Press.

\medskip

\noindent Bailin, J., \& Steinmetz, M. 2005, ApJ, 627, 647

\medskip

\noindent Banerjee, A., \& Jog, C.J. 2011, ApJ, 732, L8

\medskip

\noindent Bett, P., Eke, V., Frenk, C.S., Jenkins, A., Helly, J., \& Navarro, J. 2007, MNRAS, 376, 215

\medskip

\noindent Binney, J., Jiang, I.-G., \& Dutta 1998, MNRAS, 297, 1237

\medskip

\noindent Binney, J., \& Tremaine, S. 1987, Galactic Dynamics, Princeton: Princeton Univ. Press

\medskip

\noindent Buote, D.A., \& Canizares, C.R. 1996, ApJ, 457, 565

\medskip

\noindent Chandrasekhar, S. 1969, Ellipsoidal figures of equilibrium, New Haven: Dover

\medskip

\noindent Cuddeford, P. 1993, MNRAS, 262, 1076

\medskip

\noindent Das, M., \& Jog, C.J. 1995, ApJ, 451, 167

\medskip

\noindent de Zeeuw, T., \& Pfenniger, D. 1988, MNRAS, 235, 949

\medskip

\noindent Dubinski, J., \& Chakrabarti, D. 2009, ApJ, 703, 2068

\medskip

\noindent Helmi, A. 2004, ApJ, 610, L97

\medskip

\noindent Ideta, M., Hozumi, S., Tsuchiya, T., \& Takizawa, M.
2000, MNRAS, 311, 733

\medskip

\noindent Jing, Y.P., \& Suto, Y. 2002, ApJ, 574, 538

\medskip

\noindent Kuijken, K., \& Gilmore, G. 1989, MNRAS, 239, 571

\medskip

\noindent Mera, D., Chabrier, G., \& Schaeffer, R. 1998, A \& A, 330, 953

\medskip

\noindent Pranav, P., \& Jog, C.J. 2010, MNRAS, 406, 576

\medskip

\noindent Rohlfs, K. 1977, Lectures on density wave theory, Springer-Verlag: Berlin

\medskip

\noindent Ryden, B.S. 1990, MNRAS, 244, 341

\medskip

\noindent Saha, K., \& Jog, C.J., 2006, MNRAS, 367, 1297

\medskip

\noindent Sackett, P.D., \& Sparke, L.S. 1990, ApJ, 361, 408

\medskip

\noindent Sparke, L.S. 1994, ApJ, 280, 117

\medskip

\noindent Vera-Ciro, C.A., Sales, L.V., Helmi, A., Frenk, C.S., Navarro, J.F., Springel, V., Vogelsberger, M., White, S. D. M. 2011, MNRAS, 416, 1377

\bigskip

\noindent {\bf Appendix A : Potential and force inside a prolate shell within dissimilar surfaces}

In this paper we treat spheroids of radially varying eccentricity. In analogy with a homoeoid which is a shell bounded by two similar spheroids, we define a heteroid to be a shell bounded by two dissimilar spheroids (Section 2.2).
 We show next that the results from the standard homoeoid potential theory, namely the constancy 
of potential inside a shell and no net force within it, are good approximations even for the case of a shell with dissimilar surfaces - when the radial variation in eccentricity is small as seen in realistic systems. 
This is an important general result and to our knowledge the heteroid case has not been studied systematically 
in the past. 

\medskip

\noindent {\it Potential inside a shell within dissimilar spheroids:}
Recall from Section 2.2 that
for the dissimilar prolate surfaces with the variation in eccentricity, the
dimensionless scale factor $S$ (see eq.[21]) denotes the deviation from the case of constant internal potential:
$$S = \left[\frac{1+\sigma \,cos^2v}{1+ (\sigma/3)}\right] \eqno(A1) $$
\noindent where $\sigma$ denotes the dimensionless change in the eccentricity with radius. 

In Fig. A1, we plot the values of this scale factor versus the major axis $a/a_c$ normalized w.r.t. the core radius,
for decreasing $q$ (as given by eq. [41]) with $\alpha=-0.1$ for the prolate co-ordinate $v$= 0 and $v = \pi/2$
and similarly for increasing prolateness with $q$ increasing (as given by eq. [41]) with $\alpha=0.1$.
This figure clearly shows that 
the scale factor values are small ($<$ a few \%). Hence the deviation from the case of similar shells treated in homoeoidal theory is non-zero but small.
\begin{figure}
\begin{center}
\includegraphics[height=6.0cm,width=6.5cm]{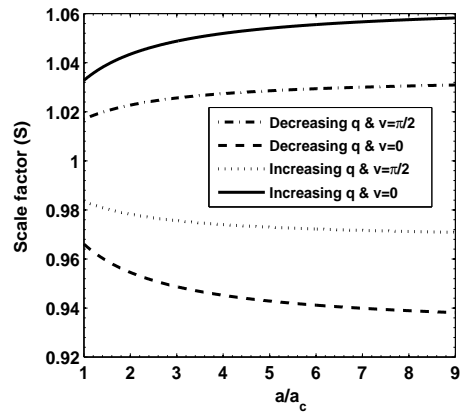}
\end{center}
\caption {The values of the dimensionless scale factor, S, that denote the deviation of the surface density within dissimilar surfaces compared to the case of similar surfaces vs.the dimensionless major axis, for increasing and decreasing prolateness, and for different values of the prolate co-ordinate $v$. Note that the deviation from the similar shells corresponding to the usual homoeoidal case is small $<$ a few \%.}
\end{figure}
Thus, interestingly, we find that for the small variation in 
eccentricity with radius and for the range of $q$ values as considered in this paper- the assumption 
of a constant potential within a shell is a good approximation
even for the case of a shell bounded by dissimilar surfaces.  

\medskip

\noindent {\it Force inside a shell within dissimilar spheroids}:
From the homoeoidal theorem we know that the force inside a shell bounded by similar spheroids is zero (e.g., Chandrasekhar 1969, Binney \& Tremaine 1987).
Here we find out the force due to a shell bounded by two dissimilar spheroids (a heteroid).
The expression of  the internal potential of  homogenous spheroid is (Binney \& Tremaine 1987, eq.2.128):
$$ \Phi (R,z) =-\pi G \rho(I a -A_1 R^2-A_3 z^2) \eqno(A2) $$
where $I, A_1 $ and$A_3$ are defined for the prolate case as follows (see Binney \& Tremaine 1987, Tables 2-1, 2-2) :
$$I = \frac{1}{e} log \left( \frac{1+e}{1-e}\right) \eqno(A3)$$
$$A_1 = \frac{1-e^2}{e^2}\left[ \frac{1}{1-e^2}-\frac{1}{2e}log \left( \frac{1+e}{1-e}\right)\right] \eqno(A4)$$
$$A_3 = 2 \frac{1-e^2}{e^2}\left[ \frac{1}{2e} log \left(\frac{1+e}{1-e}\right)-1 \right] \eqno(A5)$$
\noindent Hence the radial component of the gravitational field 
generated by 
a homogenous spheroid at an internal point (R,z) is:
$$
 F_R=- 2 \pi G \rho A_1 R  \eqno(A6)$$
The vertical force component is:
$$ F_z=- 2 \pi G \rho A_3 z \eqno(A7)$$

We can think of the force at an internal point of a heteroid shell as being obtained by the subtraction of force due to the outer spheroid and that due to the inner spheroid. Thus,
 the differential radial force due to the shell is:
$$ F_R=-2 \pi G \rho R (A_1^o-A_1^i) \eqno(A8)$$
\noindent where the superscripts $o$ and $i$ denote the outer and inner spheroids respectively, and where $\rho$ is the density of the shell. Similarly the vertical force is:

$$ F_z=- 2 \pi G \rho z (A_3^o-A_3^i) \eqno(A9)$$

First, check that the force inside of such a shell would be identically equal to zero
if the surfaces are similar (homoeoids). Next, we can find out the total differential force at a point 
(R,z) due to outer dissimilar
shells by adding the contributions of each shell. The magnitude of the differential
 force depends on the eccentricities of ellipsoids which surround the shell. The total differential force at a point is obtained for
 the density distribution (eq.[40]) and the eccentricities (eq.[41]) used in our work, and this is  normalised with the force due to a homogenous spheroid which 
has the surface that is passing through the point (R,z) and has that value of constant density. For normalization, a constant density ellipsoid is assumed for simplicity. The increasing $q$ (more prolate at larger radii) corresponds to $\alpha =0.1$ while decreasing $q$ (less prolate at smaller radii) has $\alpha=-0.1$ (see Section 3).
These correspond to a variation in the prolate axis ratio, $q$ = 1.5 to 1.9, and 1.5 to 1.2 respectively (Section 3). The increasing case covers the range of values of $q$ similar to the range deduced from flaring of HI data in the outer Milky Way (Banerjee \& Jog 2011, also see Section 4.2).

The values  for different semi-minor axis are given in Table A1. 
\begin{table}
\begin{center}
\caption {Normalized differential force magnitudes due to a shell within dissimilar surfaces}
\begin{tabular}{lrc}
    Increasing $q$\\
$a/a_c$ & $F_R$ & $F_z$ \\
\hline
1 & 0.0120 & 0.0018\\
3 & 0.0098& 0.0072 \\
8 & 0.0023& 0.0085\\
\hline
\end{tabular}\\
\begin{tabular}{lrc}
\\

    Decreasing $q$\\
$a/a_c$ & $F_R$ & $F_z$ \\
\hline
1 & 0.0141& 0.0020\\
3 & 0.0128& 0.0071 \\
8 & 0.0033& 0.0072\\
\hline
\end{tabular}
\end{center}
\end{table}
Clearly, the differential force is very small (with magnitudes $ \leq$ 1 \%) for the range of semi-minor axis values that we are interested in. As expected, the differential force magnitudes are highest at inner radii but still small $\sim 1 \%$. The typical value of $a_c$, the core radius, is 5 kpc in our Galaxy (Section 4), thus the ratio $a/a_c$ = 1-8 corresponds to about 5 to 40 kpc. Therefore,we can take the internal force 
due to the spheroids of varying eccentricity to be zero for our calculation. 
This further confirms that Newton's theorem is a good assumption
even for the case of 
dissimilar ellipsoids considered here. This is an important, general result and can be applied to other realistic 
cases. For example,
this is also applicable for studies of 
dark matter halos in simulations which show an even smaller
variation in the prolate or oblate axes ratio (e.g., Jing \& Suto 2002, Vera-Ciro et al. 2011). 

Since the typical dynamical timescale in a homogenous spheroid with a surface at say 3 $a_c$ = 15 kpc is $\sim$ a few $\times 10^ 8$ yrs, hence the differential force of $< 1\%$ of the force due to the inner spheroid is not likely to affect the stability of such a distribution over 100 times the dynamical timescales or over a Hubble time.

\bigskip

\noindent {\bf Appendix B : Oblate spehroidal mass distribution with
 radially varying eccentricity}

\medskip

In this appendix, we give the oblate case with varying eccentricity for the sake of completeness, with an approach similar to that developed in this paper  for the prolate case.
This problem has been looked at earlier by Ryden (1990), who however used the method of multipoles limiting to the quadrupole approximations. Thus Ryden's results are approximate and in particular not rigorous for inner regions,
as we check by comparing with the exact calculations below. Hence we give the exact calculation below. This could be of use not just for dark matter halo but also for other systems of varying eccentricity such as bulges.

The method of derivation of  physical quantities for the case of oblate spheroidal mass distribution  is  similar to the method that is described for 
the prolate spheroidal mass distributions.
So the resulting equations of potential, forces, vertical and horizontal frequencies are given here without detailed derivations. 
The density profile, axis ratio and $\sigma$ are used as the functions of semi-major axis $a$. (Remember that in the case of prolate distribution, $a$ is the semi-minor axis). Here the axis ratio is always less than $1$.

The potential of the oblate spheroidal shell on the surface $u=u_0$ in oblate spheroidal coordinate system is given in Binney $\&$ Tremaine (1987), to be:
$$
\Phi =
-\frac{GM}{a_0\,e}\times \left\{
\begin{array}{ll}
\,arcsin\left(e\right) & \mbox{when $u<u_0$}, \\
\nonumber\\
\,arcsin\left(sech\,u\right) & \mbox{when $u\geq u_0$}
\end{array}
\right.  \eqno (B1) $$
where $u$ is defined in cylindrical coordinates as:
$$R = \Delta\,coshu\,sinv  ;\: \:
 z = \Delta\,sinhu\,cosv  \eqno (B2) $$
where $\Delta =a_0e$ is a constant, $a_0 \equiv  cosh u_0$, $e=sech u_0$
 and $M$ is the mass on the surface $u=u_0$.\\

When the axis ratio varies with radius, the constraint that the shells do not cross is again given by $\sigma (a) \geq -1$
where $\sigma(a)$ is defined by eq. (14), except that $a$ is now the semi-major axis.

We next obtain the scale factor for oblate mass distribution with varying eccentricity similar to the prolate case (eq. [
22])  to check the validity of the homoeoidal theory.
From above equation (eq.[B1]), we can calculate the surface density of an
oblate spheroidal surface using Gauss's theorem, as obtained in 
Binney \& Tremaine 1987, see eq. 2.68):
 $$
 \Sigma=\frac{M}{4\pi a^2(1-e^2sin^2v)^{1/2}}
\eqno(B3) $$
Using the equation of spheroid we can find out the 
expression of surface density of a shell bounded by two dissimilar oblate spheroidal surfaces by 
taking the projection of the mass on the inner surface of the shell
to be:
$$
 \bar{\Sigma}=\frac{\rho \delta a}{\left(R^2+\frac{z^2}{q^4}\right)^{1/2}} \left(a+\frac{z^2}{q^3}\frac{dq}{da}\right)
\eqno(B4)$$

By replacing the cylindrical co-ordinates by spheroidal co-ordinates, we get:
$$
  \bar{\Sigma}=\frac{ q \rho\delta a}{\left(1-e^2sin^2v\right)^{1/2} }\left[1+\sigma cos^2v\right]
 \eqno(B5)$$ 
Using the equation (15), we rewrite the above equation as;
 $$
 \bar{\Sigma}=\frac{\delta M \,S}{4\pi a^2 \left(1-e^2sin^2v\right)^{1/2}} \eqno(B6)$$

Check that the function $S$ defined here is exactly the scale factor
as obtained for the prolate case (eq. [22]). On comparing eqs. (B3) and (B6),
it can be seen that the
case of dissimilar spheroids deviates from the similar spheroids by this scale factor.
In this case also, the values of the scale factor $S$
would be small. Hence as 
discussed in the case of prolate mass distribution, we can still apply the homoeoidal potential theory. That is, we can substitute the expression of mass of the shell
(eq. [15]) in eq. (B1) to calculate the
potential  due to a shell which is bounded  by two dissimilar oblate spheroidal surfaces.

It can then be shown that
the contribution of the potential at a point $(R,z)$ from outer shells is (this is analogous to eq. (26)):
$$
\sum_{a>a'}\delta \Phi_{int}=-4\pi G\int_{a'}^\infty\frac{aq}{e}\left[1+\frac{\sigma(a)}{3}\right]arcsin(e)\rho(a)da
\label{eq:phi_into} \eqno (B7) $$
From eq. (B7) onwards, $e\equiv sech u_a$.

The contribution from inner shells is (this is analogous to eq. (26)):
$$\sum_{a<a'}\delta \Phi_{ext}=-4\pi G\int_{0}^{a'}\frac{aq}{e}\left[1+\frac{\sigma(a)}{3}\right]
\times \, arcsin(sechu_a'')\rho(a)\,da\eqno(B8)$$
\noindent where $u_a''$ is the label of the spheroid that is confocal with the spheroid with semi-minor axis
$a$.
We can replace $sech\,u_a''$  in terms of the cylindrical coordinates $(R,z)$ using the relation:
$$\left({ae}\right)^2=\frac{z^2{sech^2u_a''}}{1-sech^2u_a''}+R^2{sech^2u_a''} \eqno(B9)$$
The potential at $\left(R,z\right)$ is
$$\Phi\left(R,z\right)=\sum_{a>a'}\delta \Phi_{int}\,+\,\sum_{a<a'}\delta \Phi_{ext}\,  \eqno (B10)$$

\noindent The force in the radial direction is
$$ F_R=4\pi G\frac{\partial}{\partial R}\int_{0}^{a'}\frac{aq}{e}\left[1+\frac{\sigma(a)}{3}\right]arcsin(sechu_a'')\rho(a)\,da  \eqno (B11)$$
At the galactic plane, eq. (B5) reduces to:
$$ sechu_a''=\frac{a\,e}{R} \eqno(B12)$$
and $a'$ becomes $R$. Substituting  eq. (B8) into eq. (B7) and taking the derivative gives:
$$F_R=\frac{4\pi G}{R}\int_{0}^{R}a^2q\left[1+\frac{\sigma(a)}{3}\right]\frac{\rho(a)}{(R^2+a^2\left(q^2-1\right))^{1/2}
}\,da \eqno(B13) $$
Then the square of the rotation velocity is: 
$$V^2_c=4\pi G\int_{0}^{R}a^2q\left[1+\frac{\sigma(a)}{3}\right]\frac{\rho(a)}{(R^2+a^2\left(q^2-1\right))^{1/2}}\,da \eqno(B14)$$

Check that for the special case of a pseudo-isothermal density distribution (eq. [40]), and for a constant axis ratio so that $q(a)=q$, the above equation reduces to

$$ V^2_c = 4 \pi G \rho_0 a^2_c q \large [\frac{1}{(1-q^2)^{1/2}} arcsin (1-q^2)^{1/2} $$
$$ \: \: \: \: \: \: - \frac {a_c}{[R^2 -a^2_c (q^2 - 1)]^{1/2}}  arctan \frac{[R^2 - a^2_c (q^2 - 1)]^{1/2}}{a_c q} \large]  \eqno (B15) $$

\noindent Hence, in the limit
$R\rightarrow\infty$, the terminal velocity $V_c$ is:

$$ V^2_c (R \rightarrow\infty) = \frac {4 \pi G \rho_0 a^2_c q}{(1-q^2)^{1/2}} arcsin (1-q^2)^{1/2} \eqno (B16) $$.

\noindent Note that this agrees with the expression obtained by Sackett \& Sparke (1990) for an oblate spheroid of constant shape, as expected.

The square of the angular frequency, $\Omega_h^2={F_R}/{R}$ is:
$$\Omega_h^2=\frac{4\pi G}{R^2}\int_{0}^{R}a^2q\left[1+\frac{\sigma(a)}{3}\right]\frac{\rho(a)}{(R^2+a^2\left(q^2-1\right))^{1/2}}\,da \eqno(B17)$$

The force in the vertical direction is:
$$ F_z=4\pi G\frac{\partial}{\partial z}\int_{0}^{a'}\frac{aq}{e}\left[1+\frac{\sigma\left(a\right)}{3}\right]arcsin\left(sechu_a''\right)\rho(a)\,da \eqno(B18)$$
Using eq. (B5), we get:
$$\lim_{z\rightarrow0}\frac{\partial\,sechu_a''}{\partial z}=\frac{zae}{R\left(R^2+a^2\left(q^2-1\right)\right)} \eqno(B19)$$
Taking the derivative in the vertical direction and using eq. (B15), eq. (B14) reduces to:
$$ F_z=4\pi Gz\int_{0}^{R}\,a^2q\left[1+\frac{\sigma(a)}{3}\right]\frac{\rho(a)}{\left(R^2+a^2\left(q^2-1\right)\right)^
{3/2}}\,da  \eqno(B20) $$
Then the square of the vertical frequency, $\nu_h^2={F_z}/{z}$ is: 
$$ \nu_h^2=4\pi G\int_{0}^{R}a^2q\left[1+\frac{\sigma(a)}{3}\right]\frac{\rho(a)}{\left(R^2+a^2\left(q^2-1\right)\right)^{3/2}}\,da \eqno(B21) $$
The expressions for $F_R$, $F_z$, $V_c$, $\Omega_h$ and $\nu_h$ are the same as those in the case of  prolate spheroidal mass distributions. The difference lies in the fact that $a$ is 
semi-major axis and the axis ratio $q$ is always $< 1$. In the prolate case, $a$ is semi-minor axis and the axis ratio is always $> 1$.

\bigskip

\end{document}